\documentclass[12pt]{article}
\usepackage{epsfig,a4}
\usepackage{cite}
    \advance\oddsidemargin by -1in
    \advance\evensidemargin
    by -1in
    \marginparsep
    0.4cm
    \advance\topmargin by
    -0.0in
    \makeindex

\def\GeV{\,\mbox{GeV}}

    \newcommand\la{\langle}
    \newcommand\ra{\rangle}
    \newcommand\beq{\begin{equation}}
    
    \newcommand\eeq{\end{equation}}
    \newcommand\beqn{\begin{eqnarray}}
    \newcommand\eeqn{\end{eqnarray}}

\begin{document}
\vspace*{3cm}

\begin{center}

{\Large \bf Cronin Effect in Drell-Yan Reaction}

\vspace{1cm}
{\large  M.B. Johnson$^{1}$, B.Z. Kopeliovich$^{2,3}$ and Ivan
Schmidt$^2$}
\\[1cm]
$^{1}${\sl Los Alamos National Laboratory,
Los Alamos, NM 87545, USA}
\\[0.2cm]
$^2${\sl Departamento de F\'isica, Universidad T\'ecnica Federico Santa
Mar\'ia,}
{\centerline {Casilla 110-V,
Valpara\'\i so, Chile}}
 \\[0.2cm]
$^{3}${\sl
Joint Institute for Nuclear Research, Dubna,
141980 Moscow Region, Russia}
\end{center}

\vspace{1cm}

\begin{abstract}

We explore the mechanism of transverse momentum broadening of fast quarks
propagating in nuclei, using Drell-Yan (DY) transverse momentum distributions
measured in the experiment E866 at FermiLab with beams of $800$ GeV protons.
Our theoretical analysis is based on the color dipole approach in the target
rest frame, which has provided a successful phenomenological description of a
variety of hadronic reactions.  The present application is relevant to the
regime of short coherence length (SCL), where the spatial extent of the
fluctuations of the projectile responsible for the Drell-Yan reaction is
short compared to the internucleon spacing.  In this limit, momentum
broadening comes from initial state interactions and is described as color
filtering, {\it i.e.} absorption of large-size dipoles leading to diminished
transverse separation and hence enhanced transverse momentum.  The
predictions we present are in good agreement with the E866 data.  The
interactions leading to the acquisition of transverse momentum arise from the
color-dipole cross section determined previously from deep-inelastic
scattering on proton targets.  Aside from the determination of the
color-dipole cross section, no other phenomenological input is needed to
explain the experimental results. The mean-square momentum broadening of
dileptons determined in a recent separate analysis of the data is likewise
well described by our theory.  These results confirm within the model studied that the origin of
momentum broadening in DY is the color dipole cross section mediating soft
initial state interactions between the parton of the projectile that
initiates the reaction and the nucleons of the nucleus, as provided by the
color dipole description. Predictions for broadening observables at RHIC are
presented.

\end{abstract}


\newpage

\section{Introduction}

Experimental facilities capable of studying collisions of heavy nuclei
at ultra relativistic energies, such as the Relativestic Heavy Ion
Collider (RHIC) at BNL and the future Large Hadron Collider (LHC) at
CERN, have been motivated by the expectation that data from such
collisions will reveal new properties of matter under extreme
conditions of temperature and density.  In recent years,
interpretations of data indeed suggest that during the early stages of
these collisions temperatures and densities are created capable of
supporting the quark-gluon plasma (QGP), a state of matter never before
produced in the laboratory and one of the more dramatic predictions of
quantum chromodynamics (QCD), the theory of the strong interactions.  
For this reason, there is great interest in these experiments and in
the challenging theoretical problem of determining properties of the
equation of state from the data.  These challenges derive in part from
the difficulties in unfolding the equation of state from uncertainties
in describing the reaction itself.  To help resolve these difficulties,
simpler nucleon-nucleus and deuteron-nucleus experiments, where the QGP
is not produced, have been performed under conditions of energy and
momentum transfer similar to those encountered in the nucleus-nucleus
collisions.

Some of the most important issues that can be resolved in
nucleon-nucleus reactions are related to parton propagation, for
example, deciding how quarks and gluons loose energy and acquire
transverse momentum as they traverse nuclei under various conditions of
coherence. The issues of parton propagation  can be cleanly studied in Drell-Yan (DY) reactions in which lepton pairs produced with hadronic beams are
detected under controlled kinematic conditions.  The important characteristic controlling nuclear effects in the DY
reaction is the coherence length, or equivalently  the coherence time, which is
the lifetime of the hard fluctuation containing the lepton pair. In the regime of short coherence length (SCL), which is of particular interest in this paper, the hard reaction occurs inside the nucleus over a short time scale, thus enabling one to identify the stage of initial state interactions , {\it i.e.} those preceding the hard interaction.   Thus, the DY reaction in the SCL is clean in the sense that the produced lepton pair hardly interacts at all on its way out
of the nucleus, conveying undisturbed information
about how the parton of the projectile that initiates the reaction
propagates in the nucleus before the production of the lepton pair
occurs.  The coherence length increases with energy, and in the regime of
long coherence length (LCL) one can no longer separate  the stages of
initial and final state interactions. In this case the whole nucleus
acts as a single scattering center. 

Notice that only in the LCL regime one can rely on QCD factorization
and translate multiple interactions of the projectile partons in the
nuclear target into a modification of the nuclear parton distribution
function (PDF). In the SCL regime the initial state interactions are 
known \cite{bbl} to break factorization.

To be useful, it is necessary to have good statistics under
the appropriate kinematical conditions. FermiLab data in $p+A$
collisions~\cite{alde,alde1,unp,vasiliev} are useful in these regards
and have been analyzed \cite{vasiliev,mbj,mmp,jklmps} as a source of
information about both quark energy loss and transverse momentum
broadening in nuclei.

In this paper we are particularly interested in momentum broadening of
quarks propagating in nuclei. Theoretical frameworks that have been
used to examine momentum broadening in various contexts include the
parton model~\cite{lqs,gqz,ww}, perturbative QCD
factorization~\cite{vitev}, diagrammatic multiple
scattering~\cite{bdmps}, and the color-dipole
approach~\cite{dhk,jkt,at,bd}. All these theories depend on a
phenomenological characterization of the strong interaction to account
for non perturbative physics that at present can not be calculated
directly from QCD. Accordingly, most require a many-body transport
coefficient that is fitted to nuclear scattering data.  An exception is
the color-dipole approach, where the non perturbative input is fixed at
the two-body level by deep-inelastic scattering on protons.

Fixing the the non perturbative input at the two-body level is
advantageous when the theoretical formulation is able to use this
information as the basic building block for predictions of
projectile-nucleus scattering.  This is precisely what is done in the
color-dipole approach, where nuclear reactions are calculated, using
Glauber-Gribov theory~\cite{glauber,gribov} and its
generalizations~\cite{krt}, in terms of a color-dipole cross section
fit to the deep inelastic data.  Consequently, once confirmed under
more highly controlled experimental conditions, the color-dipole theory
can be applied with a high level of confidence in more complicated
situations where the transport coefficients may not be able to be
determined because the required nuclear data is unavailable. Other
desirable features of the color-dipole approach include the fact that
since it does not entail a twist expansion, all higher twist terms are
naturally incorporated.  Another is that it does not need to be
corrected for next-to-leading order effects since gluon radiation is
already included in the dipole cross section.

The most important advantage of the color-dipole approach is that
nuclear effects are predicted rather than fitted to data.  In this work
we will implement its description of momentum broadening as developed
in Ref.~\cite{jkt}, and compare its predictions to the DY data
mentioned above, with specific interest in refining our understanding
of how partons propagate in nuclei. The data consist of ratios of
differential transverse momentum distributions $\sigma^{pA}_{DY}(p_T)$
for a variety of nuclei~\cite{alde,alde1,unp,vasiliev},
 \beq
R^{A/A'}(p_T)=\frac{\sigma^{pA}_{DY}(p_T)}{ \sigma^{pA'}_{DY}(p_T)}~,
\label{R2}
 \eeq
 where $p_T$ is the transverse momentum of the detected DY dilepton
pair.

The $A$-dependence of the mean-square transverse momentum 
$\Delta\left< p_T^2\right>$
 \beq
\Delta\langle p_T^2\rangle =\langle p_T^2\rangle ^A-
\langle p_T^2\rangle ^N
\label{broad}
    \eeq
 of the dilepton pair has been of special interest in previous studies
of parton propagation. Extraction of the $A$-dependence of
$\Delta\left< p_T^2\right>$ was attempted in Ref.~\cite{mmp}, using the
momentum distribution ratios from the E772 experiment~\cite{alde}.  
Unfortunately, $\Delta\left< p_T^2\right>$ can not be uniquely
determined from the ratios, requiring instead the individual cross
sections appearing in Eq.~(\ref{R2}).  A number of puzzles have arisen
from the results of this analysis, including an apparent inconsistency
with theoretical expectations~\cite{r3,jklmps}.  The situation remains
unresolved, but a reanalysis~\cite{jklmps} of the E772 data along with
more recent E866 data~\cite{vasiliev,webb} shows that the experimental
errors on $\Delta\left< p_T^2\right>$ were underestimated in the
original analysis, suggesting that there may in fact be no
contradiction between theory and experiment. Because of this impasse,
it is particularly interesting to ask whether the theory describing
these momentum distribution ratios agrees with the phenomenological
value of $\Delta\left< p_T^2\right>$ that has been determined from
them.  Thus, an additional goal of our work is to determine the
$\Delta\left<p_T^2\right>$ corresponding to the momentum distributions
predicted in the color-dipole approach and to compare them to the
recent phenomenological analysis given in Ref.~\cite{jklmps}.

Our paper is organized as follows.  In Sect. 2 we introduce the
color-dipole model applied to the transverse momentum distributions for
the DY reaction on individual hadrons.  Section 3 discusses the
extension of the theory to DY reactions on nuclei and the way in which
the nuclear medium modifies these distributions.  Section 3.1 addresses
the important issue of coherence length, and Sects. 3.2 presents the
theory and its implementation as applied to DY on nuclei.  Section 3.3
gives numerical results for the ratios of momentum distributions in
comparison to experiment, and Sect. 3.4 gives the corresponding results
for $\Delta\left< p_T^2\right>$.  Finally, in Sect. 3.5, we present
predictions for RHIC.  Various technical details justifying the
approximations made when evaluating the theory are relegated to the
Appendix.  Section 4 summarizes the paper and presents our conclusions.

\section{Drell-Yan reaction on a nucleon in the target rest frame}

\begin{figure}[tbh]
\includegraphics{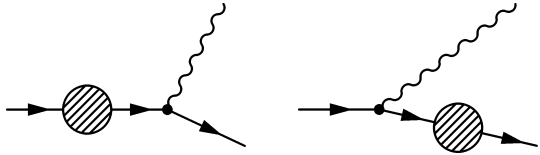}
\begin{center}
\vspace{8cm}
\parbox{13.3cm}
{\caption[dyfig]
    {\sl In the target rest frame, DY dilepton production looks like
bremsstrahlung.  A quark or an anti-quark from the projectile hadron
scatters off the target color field (denoted by the shaded circles) and
radiates a massive photon, which decays into the lepton pair.  The photon
decay is not shown.  The photon can be radiated before or after the quark
scatters.}
    \label{dyfig}}
\end{center}
    \end{figure}

The color dipole approach in the target rest frame provides an
attractive means for interpreting DY data, for reasons discussed in the
Introduction. The color dipole approach was originally proposed in
\cite{zkl} for hadronic interactions, and it was applied to
deep-inelastic scattering (DIS) in \cite{nz}. Extension to Drell-Yan
was developed by Kopeliovich~\cite{k} and subsequently by Brodsky et
al.~\cite{bhq}. According to this picture, the Drell-Yan reaction
occurs in the target rest frame when a fast projectile quark scatters
off the gluonic field of the target, radiating a massive photon
$\gamma^*$ of mass $M$ as shown in Fig.~\ref{dyfig}. In this frame, the
$\gamma^*$ is a constituent of the projectile fluctuations, which are
``frozen" by time dilation for a length of time $t_c$, the
coherence time, given by the uncertainty relation,
    \beq
t_c=\frac{2E_q}{M_{q\bar ll}^2-m_q^2}~,
    \label{coherence1}
    \eeq
where $E_q$ and $m_q$ refer to the energy and mass of the projectile
quark and $M_{q\bar ll}^2$ is the square of the effective mass of the
fluctuation,
    \beq
M_{q\bar ll}^2=\frac{m_q^2}{\alpha}+
\frac{M_{\bar ll}^2}{1-\alpha}+
\frac{k_T^2}{\alpha(1-\alpha)},
    \label{coherence2}
    \eeq
with $\alpha$ the fraction of the light-cone momentum of the incident
quark carried by the lepton pair and $k_T$ the transverse momentum of
the lepton pair.  The DY reaction occurs when the $\gamma^*$ of the fluctuation is
released by an interaction between one of the constituents of the
fluctuation and a target nucleon. Subsequently the $\gamma^*$ decays
into the observed Drell-Yan
dilepton pair $\bar ll$.

The DY cross section on a nucleon is described by
    \beq
M^2 \frac{d^2\sigma_{DY}}{dM^2dx_1}=\frac{\alpha_{em}}{6\pi
x_1}\int_{x_1}^1dx_qF_q^h(x_q)\int
d^2\rho |\Psi(\alpha ,\rho )|^2\sigma_{\bar qq}(\alpha \rho ),
\label{sigdy}
    \eeq
where $\sigma_{\bar qq}(r)$ is the universal color-dipole cross
section, $\rho$ is the
transverse distance between the $\gamma^*$ and quark in the fluctuation,
and $\Psi$ represents the light-cone distribution amplitude for the
incident quark to fluctuate
into a quark and the $\gamma^*$. In DY, the color dipole consists of the
quark before and after the release of the $\gamma^*$, whose impact
parameters differ by $\alpha\rho$. In Eq.~(\ref{sigdy}) $x_q(x_1)$ is
the fraction of
light-cone momentum of the incoming hadron $h$, for us a proton, carried
by the quark (lepton pair) and $F_q^h$ is the
average quark distribution function of the incident proton,
\beq
F_q^h(x_q)=\sum_f \left[ q_f
\left( x_q \right) +
q_{\bar f}\left( x_q \right) \right].
\label{distr}
    \eeq
The light-cone momentum fraction $\alpha$ is
given in terms of these variables as $\alpha =x_1/x_q$.

Deep-inelastic scattering from nucleons at HERA (DESY) has been used to
fix models of the color-dipole cross section. The HERA data suggests
that the color dipole cross section saturates at large $r_T$, i.e.
approaches a constant value. Golec-Biernat and W\" usthoff (GW) chose a
simple saturated form for the color dipole cross section~\cite{gbw} and
found that the following expression,
 \beq
\sigma_{\bar qq}(r_T,x)=\sigma_0(1-e^{-r_T^2/R_0^2(x)})~,
\label{cdm}
    \eeq
parametrizes well these data.
For small $r_T$,
    \beq
\sigma_{\bar qq}(r_T)\approx C\,r_T^2
\label{cdz}
    \eeq where the factor $C=\sigma_0/R_0^2(x)$ depends on Bjorken $x$.
For us, the most important point is that once the color dipole cross
section has been fitted to experimental DIS data on the nucleon, the
color-dipole formalism makes predictions for Drell-Yan both in
nucleon-nucleon~\cite{rpn} and nucleon-nucleus scattering with no
further adjustment of parameters.

The GW theory is known to describe DY on a nucleon, in the color dipole
approach, only at small Bjorken $x$, and specific
calculations~\cite{rpn} for $x_2 < 0.1$ have shown it to agree well
both in the magnitude and shape with the next-to-leading order parton
model.  Although the calculations disagreed with the E772 FermiLab
data, they are in excellent agreement with data from the E866
experiment that has recently become available~\cite{webb}.

An expression similar to Eq.~(\ref{sigdy}) exists for the momentum
distribution.
    \beq
\frac{d^4\sigma_{DY}}
{dM^2\,dx_F\,d^2p_{T}}=\frac{\alpha_{em}}{3\pi M^2}
\frac{x_1}{x_1+x_2}\int_{x_1}^1\frac{d\alpha}{\alpha^2}
F_q^h\left( \frac{x_1}{\alpha}\right)
\frac{d\sigma(qN\to \gamma^*X)}
{d\ln\alpha\,d^2p_{T}}\,.
\label{eq:dylcnucl}
    \eeq
The
differential cross section for heavy-photon radiation in a quark-nucleon
collision was derived in \cite{kst},
    \beqn\nonumber\label{eq:dylcdiff}
\frac{d\sigma(qN\to \gamma^*X)}{d\ln\alpha\,d^2k_{T}}
&=&\frac{1}{(2\pi)^2}
\int d^2\rho_1d^2\rho_2\,
\exp[{\rm i}\vec k_{T}\cdot(\vec\rho_1-\vec\rho_2)]
\Psi^*_{\gamma^* q}(\alpha ,\vec\rho_1)
\Psi_{\gamma^* q}(\alpha ,\vec\rho_2)\\
&\times&
\frac{1}{2}
\Bigl[\sigma_{q\bar q}(\alpha\rho_1)
+\sigma_{q\bar q}(\alpha\rho_2)
-\sigma_{q\bar q}(\alpha|\vec\rho_1-\vec\rho_2|)\Bigr]\ .
    \eeqn
This is described in detail in Ref.~\cite{krtj}. In this case, the
transverse momentum $p_T$ of the DY pair is measured.  The DY variables
$x_1$, $M^2$, $x_2$, and $k_T$ satisfy the equations,
 \beq
x_1\,x_2=\frac{M^2+k_T^2}{s}, \label{2a} \eeq \beq x_1-x_2=x_F\ ,
 \label{2b}
 \eeq
where $x_F$ is the Feynman variable. The variables $x_1$ and $x_2$ are
interpreted in the parton model as the Bjorken variables of the
annihilating quark and antiquark.  

Since the virtual photon of Eq.~(\ref{eq:dylcdiff}) originates in a projectile fluctuations at the hard scale, the quark distribution function over which this cross section is averaged to get the Drell-Yan reaction should be taken at the same scale.  Additionally, in principle the distribution function should be of leading order since gluon radiations are included in the dipole cross section.  In practice, the cross section ratios in which we are interested are rather sensitive to the details of the choice of the quark distribution function, and for our calculations we use a simple phenomenological choice~\cite{rmp}.

\section{Medium effects in DY reactions in the target
rest frame}

When the scattering takes place on a nucleus, multiple interactions
with target nucleons can give rise to various medium effects. The coherence
length determines how these arise, and it is useful to distinguish two limiting cases in which the theory simplifes.  One is the SCL limit, reached when the coherence length $\ell_c
\equiv c \langle t_c\rangle$ becomes much smaller than the
interparticle spacing $d$, $\ell_c \ll d$ (in a heavy nucleus,
$d\approx 2 fm$).  The other is the LCL limit, reached
when $\ell_c \gg R_A$, where $R_A$ is the nuclear radius. 

One of the 
important medium effects is shadowing. In the SCL there is no
shadowing because the duration of the fluctuation is so short that the
constituents of the fluctuation have no time to multiply interact with
the medium.  In the LCL, which is applicable to reactions at the LHC
and at RHIC under certain kinematic conditions, there is maximal
shadowing.

Interactions of partons in the nuclear medium also give rise to
parton energy loss and to transverse momentum broadening, to which
ratios of nuclear cross sections $R^{A/A'}$ are sensitive.  In DY,
these effects arise in the SCL as partons from the incident hadron
undergo soft interactions with nucleons of the nucleus before
experiencing a hard interaction that immediately liberates a $\gamma^*$
from one of its short-lived fluctuations. For the LCL, the nuclear
interactions that lead to shadowing are also a source of parton
momentum broadening.  Nuclear effects giving rise to momentum
broadening and shadowing are believed to arise predominantly from the
same color dipole cross section $\sigma_{\bar qq}(r_T)$ that mediates
the DY reaction on a nucleon.

Cases
intermediate between the SCL and LCL are generally more difficult to
model. Several methods have been proposed for this.  The most strict one
is the Green function method~\cite{krt}. Another one is a simple
interpolation between the SCL and LCL limits using the square of the
longitudinal form factor $F^2_A(q_c)$ expressed in terms of the
longitudinal momentum transferred in the reaction, $q_c = 1/l_c$. The
latter method was recently applied to determine quark energy loss from
ratios of $p_T$-integrated nuclear DY cross sections, using E772/E866
FermiLab data at $E_p=800\GeV$ \footnote{In this analysis energy loss was
assumed to be independent of the parton energy in accordance with
perturbative QCD calculations. However, for soft constituent partons an
effective energy loss turns out to rise with energy \cite{knpjs}.
Nevertheless, this fact does not affect the results of the
analysis~\cite{mbj}, but only their applications at higher or lower
energies.}. The kinematics of this data corresponds, for the most part, to
the SCL regime; therefore, initial state interactions are predominantly
soft and they decouple from the hard DY reaction.

Predictions for nuclear broadening in DY reaction based on the
theory~\cite{kst} for the long coherence length limit were recently given
in Ref.~\cite{krtj}. Here we concentrate on the SCL limit for which the
theory to describe quark transverse momentum broadening has been presented
in Ref.~\cite{jkt}.

\subsection{ Coherence length vs transverse momentum}

As discussed in the previous section, there are two important limits of
coherence length, those of SCL and of LCL, and distinct mechanisms of
transverse momentum broadening apply in each of them. Experimental
access to the LCL regime requires beams of sufficiently high energy to
allow fluctuations of the projectile to be frozen by Lorentz time
dilation so that their spatial extent exceeds the nuclear diameter. We
do not have so far any DY measurements to verify the mechanism of
transverse momentum broadening at play in this limit, although
predictions for $p+A$, $d+A$, and $A+A$ collisions at RHIC at
$x_F=x_1-x_2>0.5$ and at the LHC are available~\cite{krtj}. The
experimental situation is different for the SCL limit.  The most
extensive data sets we have today are the momentum distributions of the
E772/E866 FermiLab experiments.  To better define the dividing line
between the two regimes for the momentum distributions, we examine in
this section the coherence length as a function of transverse momentum.

We proceed by calculating the coherence length of the fluctuation in
the incident proton containing the $\gamma*$ of transverse momentum
$k_T$ that eventually decays into a dilepton pair of momentum $p_T$ as
measured in the laboratory.  We proceed as in Ref~\cite{mbj}, averaging
Eqs.~(\ref{coherence1},\ref{coherence2}) over $\alpha$,
    \beq
\left\la l_c\right\ra
=\frac{1}{m_N\,x_2} \frac{\int\limits_{x_1}^{1}
d\,x_q\,F^h_q(x_q)\int d^2{k_T}
\left|\widetilde\Psi_{\bar llq}
(x_1/x_q,k_T)\right|^2\,
K^{DY}(x_1/x_q)}
{\int\limits_{x_1}^{1}
d\,x_q\,
F^h_q(x_q)\int d^2{k_T}
\left|\widetilde\Psi_{\bar llq}
(x_1/x_q,k_T)\right|^2}\ ,
\label{3}
    \eeq
where $K^{DY}(\alpha )$ is
    \beq
    K^{DY}(\alpha)
=\frac{M^2(x_1,x_2,k_T)\,(1-\alpha)}
{M^2(x_1,x_2,k_T)\,(1-\alpha)+\alpha^2\,m_q^2+k_T^2}\ .
\label{4}
    \eeq
    with
    \beq
M^2(x_1,x_2,k_T)=sx_1x_2-k_T^2
\label{8}
    \eeq
    in accord with Eq.~(\ref{2a}).

The normalization integral for the light-cone distribution amplitude is
ultra-violet divergent (for radiation of transverse photons), implying
that the vacuum fluctuations are overwhelmed by infinitely heavy
$|q\gamma^*\ra$ fluctuations. Those heavy fluctuations, however, do not
contribute to the DY cross section, since they are too small to be
resolved and distinguished from a bare quark. Therefore, a proper
weight factors in the above averaging procedure must involve the dipole
interaction cross section on top of the light-cone distribution
amplitude. The Fourier transform of such a modified light-cone
distribution amplitude $\widetilde\Psi^T_{\bar llq}$ was evaluated in
Ref.~\cite{mbj} as
    \beq
\widetilde\Psi^T_{\bar llq}(\alpha,k_T)=
2\,Z_q\,\sqrt{\alpha_{em}}\,C(s)\,\vec e\cdot\vec k_T\,
\frac{i\,\alpha^2\,\tau^2}{\pi\,(\tau^2+k_T^2)^3}\ ,
\label{6}
    \eeq
    where $\hat{e}$ is the polarization of the virtual photon and
$\tau^2=(1-\alpha)M^2(x_1,x_2,k_t)+\alpha^2m_q^2$. The calculation was
done with the simplified dipole cross section $\sigma_{\bar qq}(r_T,x)=
C(x)\,r_T^2$.

\begin{figure}[tbh]
\includegraphics{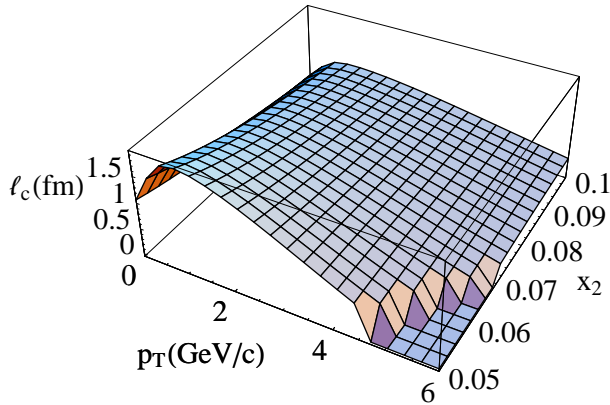}
\begin{center}
\vspace{7cm}
\parbox{13.3cm}
{\caption[dyfig]
{\sl (Color online) Coherence length for FermiLab kinematics.}
\label{clenFermilab}}
\end{center}
    \end{figure}

We show the coherence length for the FermiLab E772/E866 experiment at
$s^{1/2}=38.3\GeV$ in Fig.~\ref{clenFermilab}.  The values of $x_1$ and
$x_2$ relevant to the data for E772 are in the range $0.07<x_2<0.13$
and $0.33<x_1<0.43$~\cite{mcgaughey}.  The E866 data covers the range
$.02<x_2<.08$, with $\la x_1\ra=0.46$ and $\la M \ra=4.6$
GeV~\cite{leitch}. The values of $p_T$ cover the range from $0$ to
about $5$ GeV/c.
 Although the coherence length has been evaluated using the value of
$x_1$ corresponding to the E866 data, $\la x_1\ra=0.46$, $\ell_c$ is
rather insensitive to its precise value.  We see that the
short-coherence length limit ($\ell_c \approx 2 fm$) prevails for
$x_2>0.05$ and for all values of $p_T$ beyond $4.5\GeV/c$. The figure
is cut off when $M$ drops below $4\GeV$.  We see that this happens for
$p_T>4.5\GeV$ at $x_2=0.05$, but for $x_2>0.07$ $M<4\GeV$ throughout
the range of $p_T$ shown.

\begin{figure}[tbh]
\includegraphics{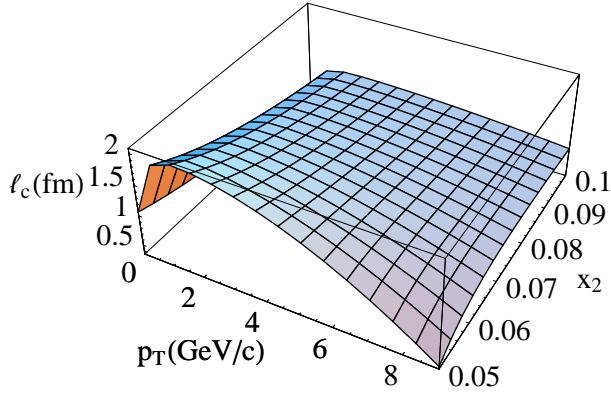}
\begin{center}
\vspace{7cm}
\parbox{13.3cm}
{\caption[dyfig]
{\sl (Color online) Coherence length for RHIC kinematics.}
\label{clenRHIC}}
\end{center}
    \end{figure}

We find the results for RHIC
kinematics shown in Fig.~\ref{clenRHIC}.  This corresponds to $x_1=x_2$
($x_F=0$) and $s^{1/2}=200\GeV$.  We see that the coherence length is
less than the internuclear spacing for $x_2>0.05$ and for all values of
$p_T$ beyond $9\GeV/c.$

\subsection{Calculating momentum distribution in the SCL limit }

Calculations of the previous section show that DY production  for $x_2>0.05$  at FermiLab in the E772/E866 experiments with 800 GeV protons and at RHIC with $s^{1/2}=200\GeV$ protons at $x_F=0$ explore momentum distributions in the SCL limit. In
this subsection we calculate these
distributions  using the theory of Ref.~\cite{jkt}.  We will compare these calculations to the E866 data in the
next subsection.  We do not show comparisons with the E772 data 
because the transverse momentum distributions are essentially 
identical to those with the more recent and reliable data from E866. 
Later on, we show predictions for transverse momentum observables 
under kinematic conditions appropriate to RHIC.

It was recognized in that work that at high energies, the multiple
interactions of a quark propagating in the nuclear medium can be
eikonalized and exponentiate into a factor containing the color dipole
cross section and the nuclear thickness function.  The physics of
momentum broadening of the quark is therefore that of color filtering,
or absorption of large dipoles leading to diminished transverse
separation with distance.  Then, the probability distribution
$W^q(k_T,s_q)=dn_q/d^2k_T$ that a valence quark arriving at a position
$(\vec B,z)$ in the nucleus $A$ will have acquired transverse momentum
$\vec k_T$ becomes~\cite{jkt}
    \beq
W^q(\vec k_T,s_q,\vec B,z) =
\frac{N}{(2\pi )^2}\int d^2b\int d^2b'\,
e^{i\vec k_T\cdot (\vec b-\vec b')}\,
e^{-\frac{b^2+b'^2}{3\la r_{cm}^2\ra}}\,
e^{-\frac{1}{2}\sigma_{\bar qq}(\vec b-\vec b',s_q)
T_A(\frac{\vec b+\vec b'}{2}+\vec B,z)},
    \label{p-distr}
    \eeq
    where $s_q$ is the square of the quark-nucleon energy in their center-of-mass, and
    \beq
N=\frac{2}{3\pi\la r_{cm}^2\ra}
\label{norm}
    \eeq
    with $\la r_{cm}^2\ra=0.79\pm 0.03~fm^2$ the mean-square charge
radius of the proton~\cite{rmsp}, and where $T_A(b,z)$ is the nuclear
thickness function,
 \beq
T_A(b,z)=\int_{-\infty}^z dz\rho_A(b,z),
\label{TA}
 \eeq
 with $\rho_A(b,z)$ the nuclear density. To obtain the transverse
momentum acquired by a quark on the nucleus, $W^{qA}(\vec k_T,s_q,z)$,
Eq.~(\ref{p-distr}) must be averaged over the nuclear density
$\rho_A(B,z)$, which entails
    \beq
W^{qA}(\vec k_T,s_q,z)\equiv \la W^{q}(\vec k_T,s_q,\vec
B,z)\ra_N=\frac{1}{A}\int d^2B
\int dz \rho_A(B,z)W^q(\vec k_T,s_q,\vec B,z)\ .
    \label{nuclearav}
    \eeq

Notice that in the regime of SCL the soft initial state interactions
and the hard reaction (DY) cross section factorize. The universal
dipole cross section, Eq.~(\ref{cdm}), fitted to low-$x$ data for
$F_2(x,Q^2)$, can be appropriately applied to the DY cross section on a
proton, but not to the soft multiple rescatterings in nuclear matter
described via Eq.~(\ref{p-distr}).  For evaluating Eq.~(\ref{p-distr})
we therefore use the KST color-dipole cross section, which covers soft
and low-$Q^2$ data \cite{kst2}. For such data Bjorken $x$ is not a
proper variable, since the smallness of $x$ does not mean high energy
if $Q^2$ is small.  The KST cross section has the same saturated shape
as Eq.~(\ref{cdm}), but with different parameters that depend
explicitly on the quark-nucleon center-of-mass energy.  For DY
with a proton beam, $s_q$ is given by $s_q=2x_1m_NE_p$, where $E_p$ is
the laboratory energy of the protons and $m_M$ is their mass. Soft
initial state interactions also lead to gluon shadowing, which we take
into account by replacing $\sigma_{\bar qq}\rightarrow
R_G(x,A)\sigma_{\bar qq}$, where $R_G(x,A)$ is the gluon shadowing
function calculated~\cite{krt} in the Green function approach from the
KST model.  See Ref.~\cite{jkt} for more discussion. We stress that our
theoretical results do not rely on any adjustable parameters.

We express the cross section, $\sigma_{DY}^{qA}(\alpha,p_T)$, for an
incident quark to produce a DY pair on a nucleus $A$ with transverse
momentum $p_T$ as the convolution of the probability $W^{qA}(k_T,s_q)$
with the momentum distribution for a quark to produce a Drell-Yan pair
on a proton, $\sigma^{qp}_{DY}(\alpha, k_T)$,
 \beq
\sigma_{DY}^{qA}(\alpha ,p_T)=
\int d^2k_T W^{qA}(k_T,s_q)
\sigma_{DY}^{qp}(\alpha,\vec p_T- \alpha \vec k_T)\ ,
\label{convolution}
    \eeq
 To obtain the momentum distribution for an incident proton, this must
be averaged over the quark distribution function of the incident
proton, as in Eq.~(\ref{eq:dylcnucl}),
    \beq
\sigma_{DY}^{pA}(p_T)= \la\sigma_{DY}^{qA}(\alpha,p_T)\ra_\alpha ~,
\label{alphaav}
    \eeq
    where we define the bracket $\la f(\alpha )\ra_\alpha$ to denote the
integral of the function $f(\alpha )$ enclosed in it over the distribution
function of a quark in the incident proton,
    \beq
\la f(\alpha )\ra_\alpha= \int_0^1 d\alpha\,dx_qF_q^h(x_q)\,
    f(\alpha )\,\delta(\alpha-x_1/x_q)\,
\label{av}
    \eeq
The combination of momenta $\vec p_T- \alpha \vec k_T$ in
Eq.~(\ref{convolution}) accounts for the fact that the DY pair
carries away only a fraction $\alpha$ of the transverse momentum
$\vec k_T$ of a
quark in the incident proton.

We denote the $\alpha$-dependence of $\sigma^{qp}_{DY}(\alpha, k_T)$
by $ g(\alpha ,k_T)$, which we take from Eqs.~ (\ref{eq:dylcnucl})
and (\ref{eq:dylcdiff}),
\beq
   g(\alpha ,k_T) = \frac{d\sigma(qN\to \gamma^*X)}{d\ln\alpha
d^2k_{T}} \left\la \frac{d\sigma(qN\to \gamma^*X)}{d\ln\alpha
d^2k_{T}}\right\ra_\alpha^{-1}~,
\label{g}
   \eeq
evaluated using Eqs.~(B.10)  - (B.13) of Ref.~\cite{krtj}.
The transverse momentum distribution $\sigma^{pp}_{DY}(k_T)$ is of 
course obtained from
$\sigma^{qp}_{DY}(\alpha, k_T)$ by convolution,
   \beq
\sigma_{DY}^{pp}(p_T)\equiv \la\sigma_{DY}^{qp}(\alpha,p_T)\ra_\alpha~,
\label{sig-p}
   \eeq
as in Eq.~(\ref{av}).  Thus,
\beq
\sigma^{qp}_{DY}(\alpha, k_T)=
\sigma^{pp}_{DY}(k_T)g(\alpha ,k_T)~,
\label{sigqpDY}
   \eeq
 and we see that $ g(\alpha,k_T) $ is closely related to the
probability $ P(\alpha,k_T)\equiv g(\alpha,k_T)  
F^h_q(x_1/\alpha)/\alpha $ that a quark of the incident proton carrying
light cone momentum fraction $\alpha$ of the detected dilepton pair
radiates the $\gamma*$ of momentum $k_T$ that decays into the pair.

Since the production of the DY pair is a hard reaction, occurring for
$M>4\GeV$, we have used the GW color-dipole cross section to calculate
$g(\alpha ,k_T)$ in Eq.~(\ref{g}).  This said, we note that the only
reason we need $g(\alpha ,k_T)$ is for the purpose of averaging over
$\alpha$, and we therefore expect the transverse momentum distributions
on nuclei to be insensitive to the details of the color dipole model
used for calculating it.

After performing averages over the nuclear density and integrating over
$\alpha$, the momentum distribution of a DY pair produced on a nucleus,
$\sigma^{pA}_{DY}(p_T)$, is given by
   \beq \sigma^{pA}_{DY}(p_T)= \int d^2k_T \la\la
W^{q}(k_T,s_q)\ra_N\sigma_{DY}^{qp}
(\alpha,\vec p_T- \alpha \vec
k_T)\ra_\alpha~, \label{momdistr}
   \eeq
   and the final expression for the nuclear ratio becomes
   \beq
R^{A/p}(p_T)=\frac{1}{\sigma_{DY}^{pp}(p_T)}
\int d^2k_T \la\la W^{q}(k_T,s_q)\ra_N\sigma_{DY}^{qp}
(\alpha,\vec p_T- \alpha
\vec k_T)\ra_\alpha~.
\label{nuclratio}
   \eeq

Using Eq.~(\ref{p-distr}) in Eq.~(\ref{convolution}) and changing
variables, we may express the transverse momentum distribution of a quark
as
   \beq
\sigma^{qA}_{DY}(\alpha,p_T)=
\frac{1}{A}\int\frac{d^2k_T}{(2\pi )^2}\int d^2r_T
e^{ik_T\cdot r_T} e^{-\frac{r_T^2}{6\la r_{cm}^2\ra}}
U(r_T)\sigma_{DY}^{qp}(\alpha ,\vec p_T-\alpha \vec k_T)
\label{sig-albar}
   \eeq
   where
   \beq
U(r_T)=N\int d^2b\int dz \int d^2R
\rho_A(\vec R-\vec b,z)e^{-\frac{2R^2}{3\la r_{cm}^2\ra}}
e^{-\frac{1}{2}\sigma_{\bar
qq}(r_T, s_q)T_A(b,z)}~. \label{U-albar}
   \eeq
   The average over the nucleus is examined in the appendix , where it is
shown that to an excellent approximation the effect of averaging
Eq.~(\ref{p-distr}) over $(B,z)$ is to replace
   \beq
T_A(b,z)\rightarrow \langle T_A\rangle /2,
\label{avT}
   \eeq
which we will refer to as the average thickness function prescription.
The average thickness function is defined by
\beq
\langle T_A\rangle=\frac{1}{A}\int d^2b T(b)^2.
\label{avT1}
   \eeq
   We may therefore obtain $\sigma^{pA}_{DY}(p_T)$ by averaging over
$\alpha$ as in Eq.~(\ref{alphaav}) with the average $T_A$ prescription,
Eq.~(\ref{avT}).

In this case, we find a numerically convenient form for $U(r_T)$ by
expanding the term $e^{-r_T^2/R_0^2}$ of Eq.~(\ref{cdm}), and thus 
obtaining,
   \beq
U(r_T)=Ae^{\frac{\sigma_0}{2}\la
T_A\ra/2}\sum_{L=0}\frac{1}{L!}
\left(\frac{\sigma_0}{2}\,\frac{\la T_A\ra}{2}\right)^L
\,e^{-Lr_T^2/R_0^2}~.
\label{expand}
   \eeq
   This leads to
\beq
\sigma^{qA}_{DY}(\alpha,p_T)=
S_0(\alpha)\sigma^{qp}_{DY}(\alpha ,p_T)+
\int\frac{d^2k_T}{(2\pi)^2}\,
S_1(\alpha ,\vec k_T,\vec p_T)\sigma^{qp}_{DY}(\alpha ,k_T)~,
   \eeq
   where
   \beqn
S_0(\alpha)&=&e^{\frac{\sigma_0}{2}\la T_A\ra/2} \label{S0} \\
S_1(\alpha ,k_T,p_T)&=&\frac{e^{\frac{\sigma_0}{2}\la T_A\ra/2}}{\alpha^2}
\sum_{L=1}\frac{(\frac{\sigma_0}{2}\la
T_A\ra/2)^L}{\frac{L}{R_0^2}+\frac{1}{6\la r_{cm}^2\ra}}\
\frac{\pi}{L!}\
e^{-\frac{(\vec k_T+\vec 
p_T)^2}{4\alpha^2}\frac{1}{\frac{L}{R_0^2}+\frac{1}{6\la
r_{cm}^2\ra}}}~,
\label{S1}
   \eeqn
 as in Eq.~(\ref{mom-distr2}) of the appendix. The $L=0$ term has been
pulled out and simplified by noting that the integrand for this term is
strongly peaked about $k_T=0$ due to the fact that $\la r_{cm}^2\ra$ is
large on the $\GeV$ scale characterizing $\sigma^{pp}_{DY}(p_T)$.
Notice that in the KST parametrization the parameter $\sigma_0\equiv
\sigma_0(s)$ is energy dependent  and thus gets its dependence on
$\alpha$ according to the convolution in (\ref{convolution}). This
dependence, however, has a very little impact on $\la S_0\ra_\alpha$.

Because $g(\alpha,k_T)$ is strongly peaked around $\alpha=1$, the
average $\left< S_1(\alpha,\vec k_T,\vec p_T)\right>_\alpha$ is
dominated by the mean value of $\alpha$ contributing to the DY
transverse momentum distribution on a nucleon.  We call this average
$\bar \alpha_1(k_T)$, defined as
 \beq
\bar \alpha_1(k_T)\approx \sqrt{\la \alpha^2 g(\alpha ,k_T) \ra_\alpha}.
\label{alphabar1}
 \eeq
 Physically, $\bar \alpha_1(k_T)$ is the average light cone momentum 
fraction of the DY dilepton pair carried by the parton of the 
incident hadron that radiates the $\gamma*$ of the dilepton pair. 
Note that this average depends not only on the momentum $k_T$ of the 
$\gamma*$ but also on $x_1$.  Alternatively, we could define 
$\bar\alpha_1(k_T)$ to be the solution of the equation
\beq
S_1(\bar\alpha_1,k_T,p_T)\equiv \la S_1(\alpha ,\vec k_T,\vec p_T) 
g(\alpha ,k_T)
\ra_\alpha~.
\label{S1bar}
   \eeq
In practice there is only a percent or two difference between $\bar 
\alpha_1$ in the two definitions.  Because $S_1(\alpha ,\vec k_T,\vec 
p_T)$ is peaked about $k_T=p_T$, the same mean value of $\bar 
\alpha_1(k_T)$ characterizes, to a good approximation, the transverse 
momentum distribution both on the nucleon and the nucleus. 
Accordingly $\bar \alpha_1(k_T)$ is found to be essentially
independent of A.  We also find that for a given $x_1$ and $M$, 
$\bar\alpha_1(k_T)$ depends smoothly on $k_T$.

\begin{figure}[tbh]
\includegraphics{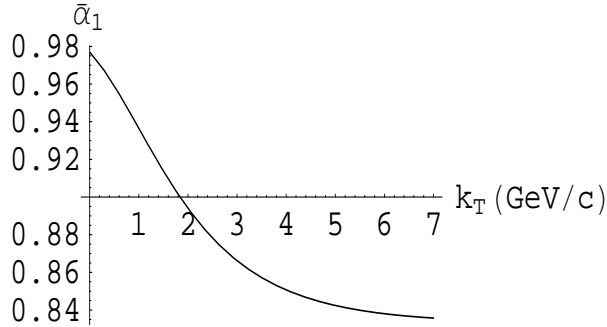}
   \begin{center} \vspace{7cm} \parbox{13.3cm} {\caption[alphac]
{\sl The quantity $\bar\alpha_1(k_T)$ of Eq.~(\ref{S1bar}) for $800\GeV$
protons incident on $^{184}W$.}
   \label{alphaf}}
   \end{center}
   \end{figure}

We show $\bar\alpha_1(k_T)$ for $^{184}W$ in Fig.~\ref{alphaf} for
$x_1=0.46$, corresponding to E866.
 The quantity $\bar\alpha_1(k_T)$ shown there can be fitted by the
function
 \beq
\bar\alpha_1(k_T)= \bar\alpha_1(0)+(ak_T+bk_T^2)/(1+ck_T^2)~.
\label{parmet}
 \eeq
 The parameters in Eq.~(\ref{parmet}) are $\bar\alpha_1(0)=0.977$,
$a=-0.0281$, $b=-0.0178$, and $c=0.134$. This function is relatively
insensitive to $x_2$ over the range encompassed by the data.  However,
at larger $k_T$, $g(\alpha,k_T)$ spreads out sufficiently in $\alpha$
for $\bar \alpha_1(k_T)$ to acquire some sensitivity to the interval
over which $\alpha$ is averaged in Eq.~(\ref{alphabar1}).  
Consequently, for larger $x_1$, $\bar \alpha_1(k_T)$ levels off at a
smaller value.

Because of the weak dependence of $\sigma_0$ on $s_q$, $\la S_0(\alpha
)\ra_\alpha$ in nearly independent of $p_T$.  With $\la S_0(\alpha
)\ra_\alpha$ and $\bar\alpha_1(k_T)$ so determined,
our final result for the momentum distribution is given by
 \beq
\sigma^{pA}_{DY}(p_T)=\la\sigma^{qA}_{DY}(\alpha ,p_T)\ra_\alpha=
\la S_0(\alpha)\ra_\alpha\sigma^{pp}_{DY}(p_T) +\int\frac{d^2k_T}{(2\pi)^2}
S_1(\bar\alpha_1(k_T),\vec k_T,\vec p_T)\sigma^{pp}_{DY}(k_T).
\label{final}
 \eeq
 In the next section we evaluate Eq.~(\ref{final}) and compare it to
the experimental results.

\subsection{Comparison to FermiLab E866 DY momentum distributions}

The theoretical calculations of this section correspond to the
transverse momentum distribution $\sigma^{pA}_{DY}(p_T)$ given in
Eq.~(\ref{momdistr}).  The distributions are determined once the
color dipole cross section, the nuclear density, and the gluon
shadowing corrections are specified. 

The gluon shadowing correction $R_G(x,A)$ is given in Fig.~4 of
Ref.~\cite{jkt} as a function of $x$ for several values of $A$. In this
paper we determine $R_G(x,A)$ for the nuclei of interest by
extrapolating from the values given there.  The value of $x$ at which
we need to evaluate $R_G$ is determined by the square of the transverse
momentum $k^2$ that a quark acquires scattering off the gluon cloud of
a nucleon at quark-nucleon center of mass energy squared $s_q$ and is
given by $x=4k^2/s_q$. The average value of $k^2$ is fixed by the
position of the peak of the effective gluon density in the KST model of
the color-dipole cross section at $s_q$. For $s_q=500$ GeV$^2$, we find
$k^2=.5$ GeV$^2$~\cite{jkt} and accordingly $x=~10^{-3}$. Although the
gluon shadowing correction grows with the size of the target, our
calculations show that the momentum distribution is rather insensitive
to $R_G(x,A)$ for the SCL kinematics considered in this paper.

The results that will be shown were evaluated using Eq.~(\ref{final}),
which was obtained from Eq.~(\ref{momdistr}) by a series of
approximations, accurate at the few percent level, and made to
facilitate the interpretation of the results and to streamline the
numerical calculation of them. This calculation entails integrating the
product of $S_1(\bar\alpha_1,p_T,k_T)$, given in Eq.~(\ref{S1}), and
$\sigma_{DY}^{pp}(k_T)$ over $k_T$ using the parametrizations of
$\bar\alpha_1(k_T)$ and $\sigma_{DY}^{pp}(k_T)$. We will see in the
next section that the transverse momentum dependence of $\bar\alpha_1
(k_T)$ and $\sigma_{DY}^{pp}(k_T)$ are key ingredients for
understanding $\Delta\langle p_T^2\rangle$.

We begin our study with the examination of the E866 FermiLab fixed
target experiment with an $800\GeV$ proton beam.  The relevant data are
nuclear distribution ratios of Eq.~(\ref{R2})  found in
Refs.~\cite{vasiliev,unp,leitch}.  The cross sections appearing in
these ratios are doubly differential distributions binned in $M^2$ (or
$x_1$) and $x_2$,
 \beq
\sigma^{pA}_{DY}(p_T)\equiv\frac{d^4\sigma^{pA}_{DY}}
{dM^2\,dx_2\,d^2k_{T}}.
    \eeq
 The individual cross sections that appear in these ratios were not
determined as accurately as the ratios and are therefore not available.  
The data is binned in unique $x_2$ intervals within a tight range in
$x_1$, {\it i.e.} $\la x_1\ra\approx 0.46$~\cite{leitch}.

Of these sets, those which have $0.1>x_2>0.05$ correspond to the
short-coherence length limit, $\la\ell_c\ra< 2 ~\rm{fm}$, where the
momentum broadening is determined by initial state interactions, and it
is this data set to which we compare our theory.  For the E866 data
corresponding to smaller $x_2$, the coherence length begins to exceed
the internucleon spacing at the center of the nucleus. The coherence
length remains at the same time smaller than the nuclear diameter for
these data sets, so these data correspond to a situation where the SCL
and LCL mechanisms are both relevant, requiring Green's function
techniques or interpolation, as discussed earlier.

We would like to avoid data for which the LCL mechanism begins to play
a role. The reason is that the mean-square transverse momentum
corresponding to the LCL mechanism is poorly defined~\cite{krtj}.
Although our recent analysis in Ref.~\cite{jklmps} found no evidence
for the mean-square momentum to depend on coherence length, the large
statistical errors do not rule out as much as a factor of two variation
among the various data sets. Thus, we restrict our study to those momentum distribution ratios that
correspond to short coherence length with $0.1>x_2>0.05$ and for which
the measured broadening of dileptons is related to soft initial state
interactions and broadening of the projectile quark.

\begin{figure}[tbh]
\includegraphics{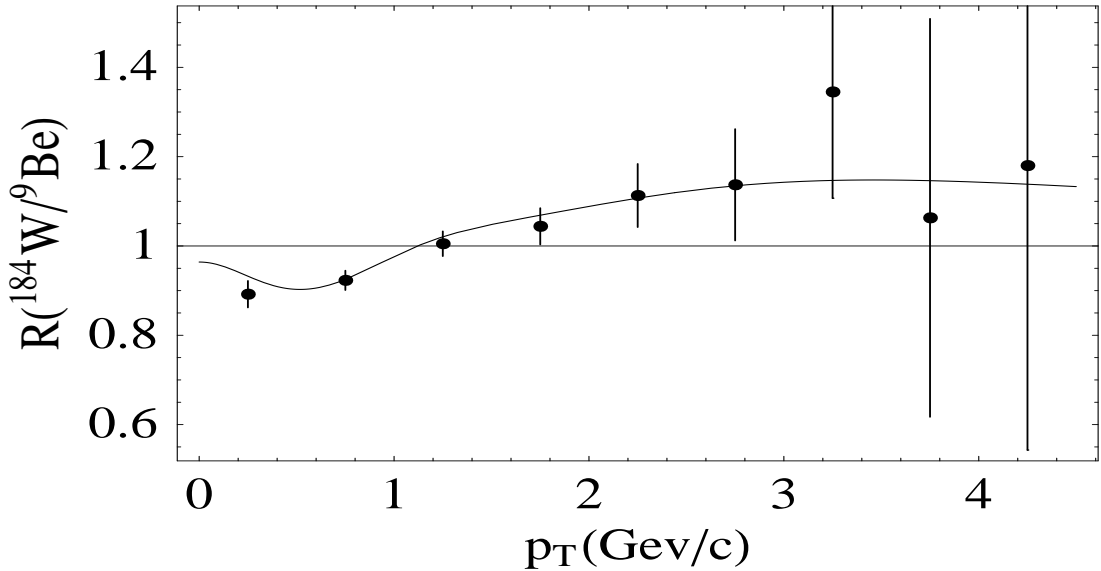}
\begin{center}
\vspace{9cm}
\parbox{13.3cm}
{\caption[tdm11c]
    {\sl Comparison of the theoretical prediction for $R^{W/Be}(p_T)$
vs. $p_T$ (in $\GeV/c$) with experiment for $x_2 = 0.05$.  Data are
from the FermiLab E772/E866 collaboration~\cite{vasiliev,unp,leitch}.}
    \label{tdm11f}}
\end{center}
    \end{figure}

For $\sigma_{DY}^{pp}(k_T)$ we have chosen a parametrization similar
to that in Ref.~\cite{knpjs},
   \beq
\sigma_{DY}^{pp}(p_T) = N\,
\frac{\left(1+\frac{p_T^2}{\lambda_1^2}\,
e^{-p_T^2/\lambda_2^2}\right)}
{(1+p_T^2/\lambda_3^2)^n}\ ,
\label{90}
   \eeq where the factor in the numerator is introduced to describe a
possible forward minimum in the cross section indicated by data
\cite{webb}. It was found that the shape of the $p_T$ dependence does
not vary with dilepton mass and $x_2$. A global fit to Drell-Yan data
from $pp$ collisions measure by the E866 experiment for different $M$
and $x_2$ bins~\cite{webb}, with common parameters $n$, $\lambda_i$ and
different normalization factors $N$, led to quite a good
description~\cite{irina}, $\chi^2/d.o.f. = 1.5$. The values of the
parameters are $\lambda_1=0.74 \pm 0.12$; $\lambda_2=0.59 \pm 0.06$;
$\lambda_3=3.03 \pm 0.25$;  $n=6.24\pm0.74$.

Notice that Eq.~(\ref{sig-p}) is not the same as
$\sigma_{DY}^{pp}(p_T)$ calculated as in Eq.~(\ref{eq:dylcdiff}). By
using the phenomenological proton-proton transverse momentum
distribution we compensate for the fact that a portion of the
transverse momentum distribution arises from the density matrix of the
initial quark in the projectile that is not yet accounted for in
Eq.~(\ref{eq:dylcdiff}). This explains why the measured DY momentum
distribution falls off more slowly with transverse momentum than
Eqs.~(\ref{eq:dylcnucl}) and (\ref{eq:dylcdiff}) evaluated as explained
in the text.

Our prediction for $R^{W/Be}(p_T)$ is shown in Fig.~\ref{tdm11f}.  For
this calculation, the gluon shadowing correction was determined to be
$R_G=0.9$ for $^{184}W$ and $R_G=1.0$ for $^9Be$. One sees that the
theory is in quite good agreement with experiment.  Notice that the
details of the Cronin effect, including the rise above $1$ for
$p_T\approx 3$ in Fig.~\ref{tdm11f}, are completely explained with no
adjustable parameters. For $W$, the value of $<T_A>=1.35$, and for $Be$
it is $<T_A>=0.288$.  We use densities from Ref.~\cite{PDT}.

\begin{figure}[tbh]
\includegraphics{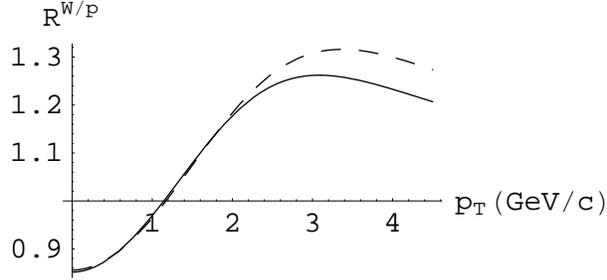}
\begin{center}
\vspace{7cm}
\parbox{13.3cm}
{\caption[comparec]
    {\sl $R^{W/p}(p_T)$ for $^{184}W$ with
$\bar\alpha_1=\bar\alpha_1(p_T)$ (solid curve), and with
$\bar\alpha_1=\bar\alpha_1(p_T=0)=0.9775$ (dashed curve).}
    \label{comparef}}
\end{center}
    \end{figure}

The importance of the momentum dependence of $\bar\alpha_1 (k_T)$ is
illustrated in Fig.~\ref{comparef}, where we compare the calculation of
$R^{W/p}$ with $\bar\alpha_1\rightarrow \bar\alpha_1(k_T=0)$ with the
same quantity, but now with $\bar\alpha_1(k_T)$. We see that retaining
the dependence on $k_T$ makes $R^{W/p}$ drop somewhat faster at larger
values of $p_T$.

\begin{figure}[tbh]
\includegraphics{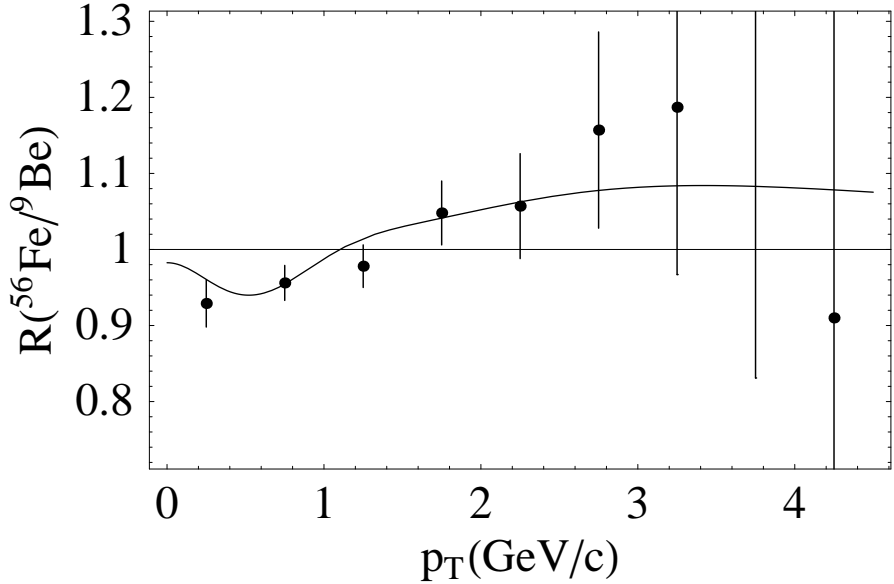}
\begin{center}
\vspace{8cm}
\parbox{13.3cm}
{\caption[tdm22c]
    {\sl Comparison of the theoretical prediction for $R^{Fe/Be}(p_T)$
vs. $p_T$ (in $\GeV/c$) with experiment for $x_2 = 0.05$.  Data are
from the FermiLab E772/E866 collaboration~\cite{vasiliev,unp,leitch}.}
    \label{tdm22f}}
\end{center}
    \end{figure}

Our prediction for $R^{Fe/Be}(p_T)$ is shown in Fig.~\ref{tdm22f}.
For the calculation, we found the gluon shadowing correction to be $R_G=0.95$
for $Fe$. We find comparable agreement for both $R^{W/Be}$ and
$R^{Fe/Be}$.

\subsection{Transverse momentum broadening}

The transverse momentum broadening for a dilepton produced on a nucleus
is defined in Eq.~(\ref{broad}) where the mean momentum is defined as,
    \beq \langle p_T^2\rangle =\frac{\int dp_T^2 p_T^2
\sigma_{DY}^{pA}(p_T)}{\int dp_T^2 \sigma_{DY}^{pA}(p_T)}.
    \label{meanp}
    \eeq
    The dependence of the broadening $\Delta\langle p_T^2\rangle$ on
$A$ for the production of heavy lepton pairs has been studied
experimentally using the E772 data in Ref.~\cite{mmp}, which was
revised recently in \cite{jklmps} with the inclusion of the E866 data.  
It was found that for large A
    \beq
\Delta\langle p_T^2\rangle^A_{exp} =  D\, (A/2)^{1/3} \GeV^2~,
\label{exp}
    \eeq
    where the factor $D$ was found to be for the two sets of data,
    \beqn
D(E772)=(0.029 \pm 0.008 +0.009)\,(\GeV/c)^2\ ;
\nonumber\\
D(E866)=(0.059 \pm 0.009 + 0.01)\,(\GeV/c)^2\ .
\label{100}
    \eeqn
 The last correction here is the systematic error, with a sign known to
be positive. This correction, if added, must be applied to both E772
and E866 simultaneously.  It comes from the theoretical uncertainty
related to the lack of data at large $p_T$.

We next calculate $\Delta\langle p_T^2\rangle $ in the theory just
described.  Because the evaluation of $\langle p_T^2\rangle $ in
Eq.~(\ref{meanp}) entails an integration over all transverse momenta,
we are lead to delta functions that require evaluations of various
quantities at $\vec r_T=0$.  For this reason, we begin by transforming
to coordinate space. From Eq.~(\ref{sig-albar}) we easily find that
    \beq
\sigma^{pA}_{DY}(p_T)= \langle 
\sigma^{qp}_{DY}(\alpha,r_T)\rangle_\alpha =\frac{1}{A}\int 
d^2r_Te^{i\vec p_T\cdot
\vec r_T}e^{-\frac{\alpha^2r_T^2}{6\la r_{cm}^2\ra}}
U(\alpha r_T) \langle \sigma^{qp}_{DY}(\alpha,r_T)\rangle_\alpha
    \label{sig1}
    \eeq
    where $U(r_T)$ is given in Eq.~(\ref{U-albar})
and
\beqn
\langle \sigma^{qp}_{DY}(\alpha,r_T)\rangle_\alpha 
\equiv\sigma^{pp}_{DY}(r_T) = \int \frac{d^2p_T}{(2\pi)^2}e^{-i\vec 
p_T\cdot \vec r_T}
\langle \sigma_{DY}^{qp}(\alpha,p_T)\rangle_\alpha~.
\label{mom3}
\eeqn
Multiplying Eq.~(\ref{sig1}) by $p_T^2$ and integrating over all 
$\vec p_T$, we find, using the fact that $\sigma_{\bar qq}(r_T)$ is 
proportional
to $r_T^2$ at small $r_T$ (see Eq.~(\ref{cdz})),
\beqn
\langle p_T^2\rangle^A&=&\langle p_T^2\rangle^N
+\frac{4CN\langle\alpha^2\rangle}{2A}\int d^2b\int dz\int
d^2R\rho (R-b,z)e^{-\frac{2R^2}{3\la r_{cm}^2\ra}}T_A(b,z)~,
\label{ptA}
\eeqn
where
\beq
\langle p_T^2
\rangle^N=-\frac{\nabla^2\sigma_{DY}^{pp}(r_T)}{\sigma_{DY}^{pp}(r_T)}|_{r_T=0}+\frac{2}{3}
\frac{\langle\alpha^2\rangle}{\langle r_{cm}^2\rangle}
\label{ptN}
\eeq
is the mean transverse momentum of DY reaction on a proton target, and where
\beq
\langle \alpha^2 \rangle
=\frac{1}{\langle\sigma^{qp}_{DY}(\alpha,r_T=0)\rangle_\alpha}\langle\alpha^2\sigma^{qp}_{DY}(\alpha,r_T=0)\rangle_\alpha 
~.
\label{al2def}
\eeq

The quantity $\langle \alpha^2\rangle$ in Eq.~(\ref{al2def}) is the
fraction of the light cone momentum carried by the parton of the
projectile that radiates the $\gamma*$, averaged over the transverse
momentum for fixed $x_1$, as can be seen by transforming back to
momentum space.  Using the Fourier transform in Eq.~(\ref{mom3}) and
the definition of $\sigma^{qp}_{DY}(\alpha, k_T)$ in
Eq.~(\ref{sigqpDY}),  with the quantity $\bar\alpha_1(p_T)$ shown in
Fig.~\ref{comparef} as defined in Eq.~(\ref{alphabar1}),
Eq.~(\ref{al2def}) may be equivalently written as
 \beq
\langle \alpha^2 \rangle =\frac{\int d^2p_T
\sigma_{DY}^{pp}(p_T)\bar\alpha_1(p_T)^2}{\int d^2p_T
\sigma_{DY}^{pp}(p_T)}~.
\label{al2fin}
 \eeq
 We find that $\langle \alpha^2\rangle\approx 0.86$.
This value is slightly smaller than the previous estimates in
Ref.~\cite{krtj}  ($\langle \alpha^2\rangle\approx 0.9$) because the
integrand in Eq.~(\ref{al2fin}) peaks at slightly higher $p_T$ due to
the fact that the phenomenological momentum distribution of the
proton $\sigma_{DY}^{pp}(p_T)$ is softer than the theoretical
distribution used in Ref.~\cite{krtj} (see discussion below
Eq.~(\ref{sig-p})).

The first term in Eq.~(\ref{ptN}) corresponds to the contribution to
the mean transverse momentum arising from the hard interaction, and the
second term is the contribution arising from the primordial transverse
momentum of a quark in the projectile hadron.  The contribution to
$\langle p_T^2\rangle^N$ arising from the hard interaction is formally
infinite.  As long as there is no shadowing, as in the SCL, this gets
subtracted to obtain $\Delta\langle p_T^2\rangle$,
 \beq
\Delta\langle p_T^2\rangle\equiv 
\langle p_T^2\rangle^A-\langle p_T^2\rangle^N
=\frac{4CN\langle\alpha^2\rangle}{2A}\int d^2b\int dz\int d^2R\rho
(\vec R-\vec b,z)e^{-\frac{2R^2}{3\la r_{cm}^2\ra}}T_A(b,z)~.
\label{delpt1}
    \eeq
This cancellation is not complete for the LCL case~\cite{krtj}, but we
have avoided the associated complications by restricting our
attention to $x_2>0.05$, where the SCL dominates.  Using
Eq.~(\ref{delpt1}) and dropping the center-of-mass density
contribution (the second term in Eq.~(\ref{delrhodef}) of the appendix),
which is only about a 3\% correction here, we find
    \beq
\Delta\langle p_T^2\rangle=\langle\alpha^2\rangle C\langle T_A\rangle~,
\label{final1}
    \eeq
 where we have used the average of $T_A$ as given in Eq.~(\ref{avT1}).  
We recall~\cite{jkt} in passing that Eq.~(\ref{final1}) for broadening
of a dilepton pair produced in DY is closely related to the broadening
experienced by a quark propagating through nuclear matter, differing
importantly by the following considerations.  First, on average a quark
in the SCL limit passes through only half the nuclear thickness before
it produces the dilepton pair (thus Eq.~(\ref{final1}) is missing the
factor of 2 that appears in Eq.~(20) of Ref.~\cite{jkt}).  Secondly,
the quark gives up only a fraction $\alpha$ of its transverse momentum
to the dilepton pair, which accounts for the extra factor of
$\langle\alpha^2\rangle$ in Eq.~(\ref{final1}).

We may now compare with the experimental result in Eq.~(\ref{exp}).
Using values of $C=4.475$ corresponding to the KST model used for our
calculations of the momentum distributions, and taking the uniform
density model for the nuclear density,
 \beq
\langle T_A\rangle = \frac{3}{2}\rho_0R_A\ ,
\label{sharpT}
\eeq
 where $\rho_0\approx 0.16 fm^{-3}$ is the central density of heavy
nuclei and $R_A\approx 1.1 A^{\frac{1}{3}}$, we find from theory that
 \beq
\Delta\langle p_T^2\rangle^A_{Th}\approx 
0.049(A/2)^{\frac{1}{3}}\GeV^2~,
\label{th}
 \eeq
 or $D=.049$.  This agrees with the central value of $D(E866)$ given in
Eq.~(\ref{100}) to within the statistical plus systematic errors and is
only slightly outside the errors on $D(E772)$.  It was concluded in
Ref.~\cite{jklmps} that large systematic errors must be assigned to the
value of $D$ in the original $E(772)$ analysis, and our results in
Eq.~(\ref{th}) are certainly consistent with this observation.

\subsection{Predictions for transverse momentum observables in the SCL
at RHIC}

Since momentum broadening in DY uniquely characterizes the momentum
broadening of the quark in the initial state, a measurement of DY at
RHIC would complement information obtained from momentum distributions
obtained from production of pions, which are additionally sensitive to
final state interactions. As seen in Fig. 3, the SCL prevails at RHIC
for DY in proton-nucleus collisions when $x_2^{RHIC}>0.05$, for a range
of transverse momentum up to several GeV/c.  This regime at
$x_F=x_1-x_2=0$ is not easily accessed experimentally with current
detectors at RHIC, but for the reason given before, it would be
interesting to measure.  The light-cone target-rest-frame formulation
provides a theoretical framework for calculating the transverse
momentum observables for RHIC, and we next use this theory for making
predictions.

In $dA$ collisions at the RHIC collider beams of deuterons are incident
on nuclei at an energy of $100$ GeV/N, corresponding to collisions
between the projectile proton with a nucleon in the nucleus at
$\sqrt{s}=200$ GeV.  In the target rest frame, this corresponds to
protons incident on the nucleus with laboratory energy of about
$E^{RHIC}_p=21$ TeV.  However, for $x_1^{RHIC}=.05$, corresponding to
the smallest value of Bjorken $x_1$ at which the SCL limit prevails,
the quarks of the beam proton that radiate the dilepton pair have an
average energy of only about $<E_q^{RHIC}>=x_1^{RHIC}E_q^{RHIC}=1.1$
TeV, which is not so different from the average quark energy in the
E866 FermiLab experiments. As indicated earlier, the FermiLab E866
experiment involves larger $x_1\approx 0.46$, which is mainly
responsible for the small differences we predict.

For RHIC, we will do our calculation for the lowest value of $x_1$ for
which the SCL prevails, $x_1=x_2=0.05$.  For this value of $x_1$, we
find that the dependence of $\bar\alpha_1(k_T)$ is given by
Eq.~(\ref{parmet})  with $\bar\alpha(0)=0.9513$ $a=0.0168$, $b=-0.137$,
and $c=0.197$, corresponding to a somewhat stronger fall-off of
$\bar\alpha_1(k_T)$ with $k_T$ than we found for E866.

For gluon shadowing, the larger RHIC energies do not lead to much
different $R_G(x,A)$ from the values found for heavy nuclei at
FermiLab.  For determining the effective value of
$x=4k^2/s_q$~\cite{jkt} for $R_G(x,A)$, we take $s_q=2mE^{RHIC}_q=2000$
GeV$^2$, corresponding to $k^2=.6$ GeV$^2$.  This gives $x=1.1~10^{-3}$
and $R_G=0.91$ from Fig. 4 of Ref.~\cite{jkt}.

\begin{figure}[tbh]
\includegraphics{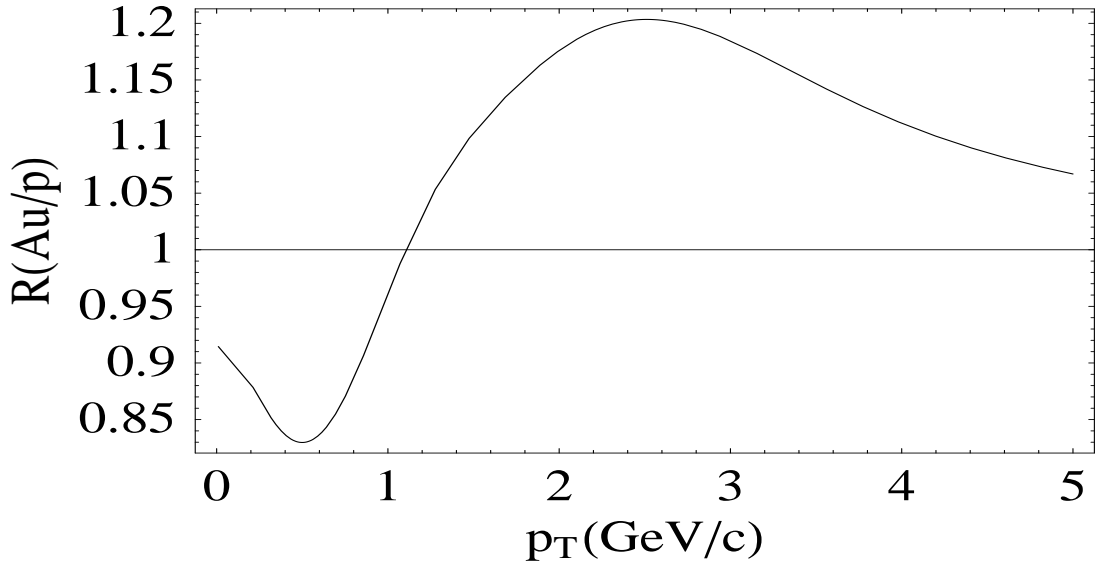}
\begin{center}
\vspace{9cm}
\parbox{13.3cm}
{\caption[tdm33c]
    {\sl Theoretical prediction of $R^{Au/p}(p_T)$ vs. $p_T$ (in
$\GeV/c$) for DY at RHIC, $\sqrt{s}=200$~GeV for $x_F=0$ ($x_2 =
0.05$). }
    \label{tdm33f}}
\end{center}
    \end{figure}

The transverse momentum distribution for protons on gold, according to
our theory, is shown in Fig~\ref{tdm33f}.  We see a clear Cronin peak
at $p_T=2$ GeV of about the same size seen in DY ratios at FermiLab and
in pion production at RHIC.  The ratio to protons was calculated
assuming that the transverse momentum dependence of DY production on
protons is the same as measured in E866 at FermiLab and given in
Eq.~(\ref{90}).  One should of course use one more appropriate for the
RHIC kinematics, but this is presently unknown.  Using the distribution
measured in E866 is reasonable since the kinematics are similar and
presumably differences will tend to cancel in taking the ratio, but
still there may be quantitative differences when the result is
calculated with a known momentum distribution on the proton.

The value of $\Delta\langle p_T^2\rangle$ corresponding to the
distribution given in Fig.~\ref{tdm33f} is given by Eq.~(\ref{final1})
with $\langle\alpha^2\rangle =0.67$, $C(s_q)=6.04$, and $\langle
T_A\rangle=1.5$.  The value of $\langle\alpha^2\rangle$ is somewhat
smaller than it was for E866 since $\bar\alpha_1(k_T)$ falls off more
quickly in $k_T$ because of the smaller $x_1$.  This decrease
(accidentally) almost completely compensates the increase in the
transverse momentum acquired by the quark before it produces the
dilepton pair, which is determined by $C(s_q)$ from the KST
color-dipole cross section at $r_T=0$ and has a weak logarithmic
increase with energy. We find
 \beq \Delta\langle
p_T^2\rangle^A_{Th}\approx 0.042(A/2)^{\frac{1}{3}}\GeV^2~,
\label{thau}
 \eeq
 corresponding to $D=0.042$, very close to our theoretical value of $D$
corresponding to E866 and given in Eq.~(\ref{th}).

\section{Summary and conclusions}
\label{conc}

The color dipole approach formulated in the target rest frame provides
a natural framework for examining nuclear modifications to parton
propagation in collisions of hadronic projectiles on nuclei. In this
paper, we are interested in transverse momentum broadening of fast
quarks propagating in nuclei, which we have examined in a particular
kinematic regime using the color dipole approach.  The color-dipole
approach has predictive power since it relates transverse momentum
observables in nuclear collisions to the color-dipole cross section
determined from deep-inelastic scattering on a nucleon target.  Once
the color-dipole cross section is determined, predictions for these
observables can be made with no parameters to be adjusted~\cite{jkt}.
Transverse momentum broadening in Drell-Yan production has been
recently measured with good accuracy in $p+A$ collisions in the E866
experiment at FermiLab~\cite{vasiliev,unp,leitch} with $E_p=800\GeV$.
In this paper, we have applied our theory to the FermiLab data.

In the short coherence length limit, $\ell_c\ll R_A$, energy loss and
momentum broadening are mediated by multiple interactions of the
incident quark with target nucleons before the emission of the
$\gamma^*$.  The SCL limit is appropriate for the FermiLab data and for
application of the theory given in Ref.~\cite{jkt}.  The transverse
momentum distributions we calculate from this theory, presented in
Sect.~3.3, are in good agreement with the measured ones. The value of
$\Delta\langle p_T^2\rangle$ determined from the same theory gives the
result presented in Eq.~(\ref{th}), which is comparable to the
empirical determination of the same quantity~\cite{mmp,jklmps}.

We have thus confirmed within the model studied that transverse momentum broadening in the SCL
arises from soft initial state interactions experienced by the parton
initiating the hard reaction, as described in the color-dipole
approach.  In addition, our results support the validity of the method
of determining the mean-square transverse momentum distribution as
given in Ref.~\cite{jklmps}.

Predictions for transverse momentum observables at RHIC in the short
coherence limit have been presented, complementing our earlier
study~\cite{krtj} of broadening in the long coherent length regime.

\section{Acknowledgements} We would like to acknowledge our
collaborators I.K.~Potashnikova, A.V.~Tarasov, J.~Raufeisen, M.~Leitch,
J.~Moss, and P.~McGaughey.  This work was supported in part by U.S.  
Department of Energy, the Research Ring "Center of Subatomic Studies"
(Chile), Fondecyt (Chile) grant 1050519, and by DFG (Germany) grant
PI182/3-1.

\newpage

\def\appendix{\par
    \setcounter{section}{0}
    \setcounter{subsection}{0}
    \def\thesection{Appendix \Alph{section}: }
    \def\thesubsection{\Alph{section}.\arabic{subsection}}
    \def\theequation{\Alph{section}.\arabic{equation}}
    \setcounter{equation}{0}}
\appendix

\section{Average over nuclear density}
\setcounter{equation}{0}

In this appendix, we are interested in computing the average over the
nuclear density in the expression for the momentum distribution of a DY
pair produced on a nucleus, $\sigma^{pA}_{DY}(p_T)$,
    \beq
\sigma^{pA}_{DY}(p_T)=\int d^2k_T
\la\la W^{q}(k_T,s_q)\ra_N\sigma_{DY}^{qp}(\alpha,\vec p_T- 
\alpha \vec k_T)\ra_\alpha.
\label{momdistr1}
    \eeq
    For the purpose of evaluating the nuclear average, we note that the
$\alpha$ dependence in $\sigma_{DY}^{qp}(\alpha,p_T)$, as given by
Eq.~(\ref{eq:dylcdiff}), is sharply peaked around $\alpha=1$.  This
means that it is justified to replace the more slowly-dependent
functions of $\alpha$ by averages.  One source of $\alpha$ dependence
is the quark energy $s_q$ that appears in $W^q(k_T,s_q)$ through the
KST color-dipole cross section.  In the target rest frame, $s_q$ is
related to the quark energy $E_q$ by $s_q=2m_pE_q$, with
$E_q=E_px_1/\alpha$.  As the color-dipole cross section depends only
weakly on $s_q$, the average is insensitive to the specific value,
which on average is about $\bar E_q\approx E_px_1$.  The remaining
$\alpha$ dependence is the explicit factor appearing in
Eq.~(\ref{momdistr1}).

Then, the momentum distribution $\sigma^{pA}_{DY}(p_T)$ may thus be
expressed, to an excellent approximation, as
    \beqn
\sigma^{pA}_{DY}(p_T)&\approx&\int d^2k_T \la W^{q}(k_T,\bar
s_q)\ra_N\la \sigma_{DY}^{qp}(\alpha,\vec p_T- \alpha \vec k_T)\ra_\alpha
\label{avconvolution}
\\
&\approx&\int d^2k_TW^{qA}(k_T,\bar
s_q)\sigma^{pp}_{DY}(\vec p_T-\bar \alpha \vec k_T) \label{avconvolution1}
    \eeqn
    where $\bar \alpha_1$ may depend on $k_T$, {\it i.e.} $\bar
\alpha_1 = \bar\alpha_1 (k_T)$ and where $\sigma_{DY}^{pp}(k_T)$ is
defined as in Eq.~(\ref{sig-p}). Using Eq.~(\ref{p-distr}) in
Eq.~(\ref{avconvolution1}) and changing variables, we may obtain
expressions for $\sigma^{pA}_{DY}(p_T)$ and $U(r_T)$ identical to those
in Eqs.~(\ref{sig-albar},\ref{U-albar}),
 \beq
\sigma^{pA}_{DY}(p_T)=\frac{1}{A}\int\frac{d^2k_T}{(2\pi )^2}\int d^2r_T
e^{i\vec k_T\cdot \vec r_T} e^{-\frac{r_T^2}{6\la r_{cm}^2\ra}}
U(r_T)\sigma_{DY}^{pp}(\vec p_T-\bar\alpha \vec k_T)
\label{sig-albar1}
 \eeq
 \beq
U(r_T)=N\int d^2b\int dz \int d^2R
\rho_A(\vec R-\vec b,z)e^{-\frac{2R^2}{3\la r_{cm}^2\ra}}
e^{-\frac{1}{2}\sigma_{\bar
qq}(r_T,\bar s_q)T_A(b,z)}~. \label{U-albar1}
 \eeq
 except that $\bar\alpha_1$ and $\bar s_q$ replace $\alpha$ and $s_q$.

In this appendix, we calculate $\sigma^{pA}_{DY}(p_T)$ in specific
models, and show that we are able to get quantitative agreement with
the exact expression, Eq.~(\ref{U-albar1}), if we make the average
thickness function prescription of Eq.~(\ref{avT}) for the model.

One model we consider is the uniform density model,
 \beq
\rho_A(b,z)=\rho_0\theta(R_0-r),
\label{sharp}
 \eeq
 where $R_0$ and $\rho_0$ are fixed to preserve $A$ and $\la T_A\ra$ of
a realistic density for the nucleus in question.  We find in practice
that when $R_0$ and $\rho_0$ are fixed in this fashion, the uniform
density model can be applied to calculate $\sigma^{pA}_{DY}$ for nuclei
as light as $^9Be$. For light nuclei, we use a gaussian density,
 \beq
\rho_A(b,z)=\rho_0e^{-(b^2+z^2)/R_0^2}~.
 \eeq
 where $R_0$ and $\rho_0$ are again fixed to preserve $A$ and $\la 
T_A\ra$
of a realistic density.

To calculate $U(r_T)$, we begin by writing
\beq
\rho_A(\vec R-\vec u,z)=\rho_A(u,z)+\delta\rho_A(\vec R,\vec u,z)
\label{delrhodef}
\eeq
and noting that $U(r_T)$ may be written

\beq
U(r_T)=U_0(r_T)+\delta U(r_T)
\label{main-cm}
\eeq
where
\beq
U_0(r_T)=\int d^2b\int_0^{T_A(b)} dT_A e^{-\frac{1}{2}\sigma_{\bar
qq}(r_T,\bar s_q)T_A}
\label{main}
\eeq
and
\beq
\delta U(r_T)=N\int d^2u\int dz \int d^2R
\delta\rho_A(\vec R,\vec u,z)e^{-\frac{2R^2}{3\la
r_{cm}^2\ra}}e^{-\frac{1}{2}\sigma_{\bar qq}(r_T,\bar s_q)T_A(u,z)}~.
\label{cm}
\eeq
We refer to Eq.~(\ref{cm}) as the center-of-mass correction.

In general, $U(r_T)$ is characterized by a set of numbers $U_L$,
\beq
U(r_T)=\sum_{L=0}U_Le^{-Lr_T^2/R_0^2}/L!
\label{expand1}
 \eeq
 obtained by expanding $e^{-r_T^2/R_0^2}$ in $\sigma_{\bar qq}(r_T,\bar
s_q)$ (see Eq.~(\ref{cdm})). With this expansion, we may do the
integrals over $r_T$ in Eq.~(\ref{U-albar1}), leading to
 \beq
\sigma^{pA}_{DY}(p_T)=\frac{1}{A}\sum_{L=0}\frac{\pi}
{\frac{L}{R_0^2}+\frac{1}{6\la
r_{cm}^2\ra}}\frac{U_L}{L!}
\int \frac{d^2k_T}{\bar\alpha_1^2(2\pi)^2}
e^{-\frac{k_T^2}{4\bar\alpha_1^2}\frac{1}{\frac{L}{R_0^2}+\frac{1}{6\la
r_{cm}^2\ra}}}
\sigma^{pp}_{DY}(\vec p_T-\vec k_T)~.
\label{mom-distr2}
 \eeq

In the uniform density model, $U_{0L}$, corresponding to the first
term in Eq.~(\ref{main-cm}), may be worked out analytically to give
\beq
U_{0L}=\frac{2\pi
R_0^2}{c^2\sigma_0}[\Gamma(L+3,c)-
\Gamma(L+3)+c^2\Gamma(L+1)-c^2\Gamma(L+1,c)]~,
\label{udm-N}
\eeq
where
\beq
c=\frac{2}{3}\sigma_0\la T_A\ra
\eeq
with $\la T_A\ra $ defined in Eq.~(\ref{avT1}).
For the gaussian density, $U_{0L}$ becomes
\beq
U_{0L}=A/c\int_0^\infty dx (\Gamma(L+1)-\Gamma(L+1,ce^{-x}))
\label{gdm-N}
\eeq
where now
\beq
c=\sigma_0\la T_A\ra~.
\eeq
The contribution $\delta U_L$ of the center-of-mass correction for
both the uniform density model and the gaussian model
have been evaluated numerically.  For the average thickness function
prescription,
$U_L$ may be read from Eq.~(\ref{expand}),
\beq
U_L=Ae^{\frac{\sigma_0}{2}\la T_A\ra/2}(\frac{\sigma_0}{2}\la
T_A\ra/2)^L~,
\eeq
with no center-of-mass correction.

We next illustrate the validity of the average thickness function
prescription of Eq.~(\ref{avT1}) for a heavy nucleus ($A=184$) in the
uniform density model and a light nucleus ($A=9$) in the gaussian
density model.  In these cases, we evaluate $\sigma^{pA}_{DY}(p_T)$ as
given in Eq.~(\ref{mom-distr2}) for $\bar s_q=600\GeV$ and
$\bar\alpha=1$, but similar results are obtained for energies relevant
at RHIC and for the full range of $\bar\alpha$ that enters into the
calculation of the nuclear momentum distribution in
Fig.~\ref{comparef}.

\begin{figure}[tbh]
\includegraphics{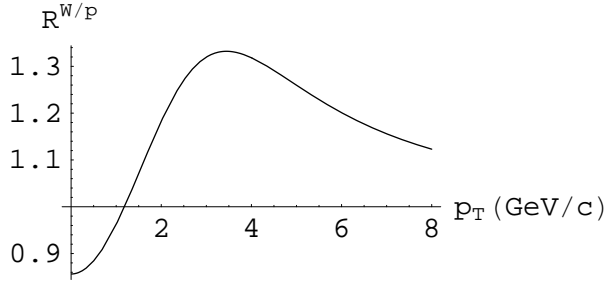}
\begin{center}
\vspace{7cm}
\parbox{13.3cm}
{\caption[distrWc]
{\sl Nuclear ratio $R^{W/p}(p_T)$ in the uniform density model
calculated using Eq.~(\ref{mom-distr2}), including the center-of-mass
correction.}
\label{distrWf}}
\end{center}
    \end{figure}

\begin{figure}[tbh]
\includegraphics{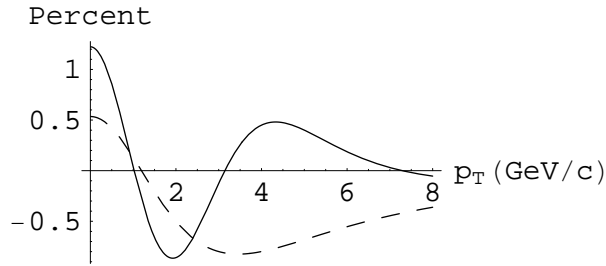}
\begin{center}
\vspace{7cm}
\parbox{13.3cm}
{\caption[delWc]
{\sl Comparison of $R^{W/p}_{UD}$ and $R^{W/p}_{ATF}$ (solid curve);
relative contribution of center-of-mass correction
to $R^{W/p}_{UD}$ in the uniform density model (dashed curve).  See
text for details.}
\label{delWf}}
\end{center}
    \end{figure}

For $A=184$, we take $\la T_A\ra=1.35$, which corresponds to a nuclear
density proportional to the charge density for $^{184}W$~\cite{PDT}.  
The parameters of the uniform density (UD) model are then $R_0=6.99~fm$
and $\rho_0=0.129~fm^{-3}$.  The result of calculating $R^{W/p}_{UD}$,
including the contributions from $U_{0L}$ of Eq.~(\ref{udm-N}) and
$\delta U_L$ calculated numerically, is shown in Fig.~\ref{distrWf}.
In Fig.~\ref{delWf}, the result in Fig.~\ref{distrWf} is compared to
the corresponding quantity calculated using
the average thickness function (ATF) prescription, $R^{W/p}_{ATF}$;
specifically, the solid curve gives
\beq
100\frac{R^{W/p}_{UD}(p_T)-R^{W/p}_{ATF}(p_T)}{R^{W/p}_{UD}(p_T)}~.
\label{UD-ATF}
 \eeq
 Clearly, $R^{W/p}_{UD}$ and $R^{W/p}_{ATF}$ are equal to about one
percent or better.  The dashed curve in Fig.~\ref{delWf} shows the
contribution of the center-of-mass correction $\delta R^{W/p}_{UD}$ to
$R^{W/p}_{UD}$; specifically, the dashed curve gives
 \beq
100\frac{\delta R^{W/p}_{UD}(p_T)}{R^{W/p}_{UD}(p_T)}~.
\label{cm-UD}
 \eeq
 The center-of-mass correction is clearly quite small, less than a 
percent.

\begin{figure}[tbh]
\includegraphics{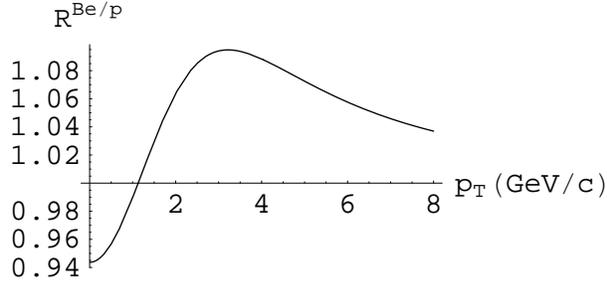}
\begin{center}
\vspace{7cm}
\parbox{13.3cm}
{\caption[distrBec]
{\sl Nuclear ratio $R^{Be/p}(p_T)$ in the gaussian density model
calculated using Eq.~(\ref{mom-distr2}), including the center-of-mass
correction.}
\label{distrBef}}
\end{center}
    \end{figure}

\begin{figure}[tbh]
\includegraphics{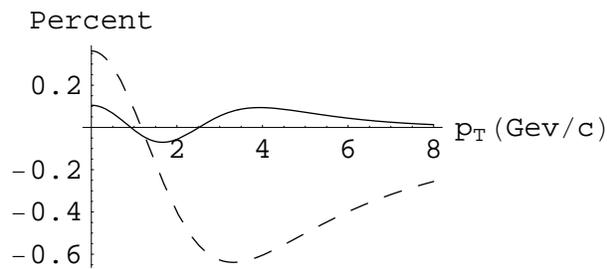}
\begin{center}
\vspace{7cm}
\parbox{13.3cm}
{\caption[delBec]
{\sl Comparison of $R^{Be/p}_{GD}$ and $R^{Be/p}_{ATF}$ (solid curve);
relative contribution of center-of-mass correction
to $R^{Be/p}_{GD}$ in the gaussian density model (dashed curve).  See
text for details.}
\label{delBef}}
\end{center}
    \end{figure}

For $A=9$, we take $\la T_A\ra=0.288$, which corresponds to a nuclear
density proportional to the charge density for $^9Be$~\cite{PDT}.  The
parameters of the gaussian density (GD) model are then $R_0=2.23~fm$
and $\rho_0=0.146~fm^{-3}$.  The result of calculating $R^{W/p}_{GD}$
including
the contributions from $U_{0L}$ of Eq.~(\ref{gdm-N}) and $\delta U_L$
calculated numerically is shown in Fig.~\ref{distrBef}.
We note in passing that essentially the same results for $^9Be$ are
found when we apply the uniform density model to this nucleus.
The solid curve
in Fig.~\ref{delBef} compares $R^{Be/p}_{GD}$ to the corresponding
quantity in the average thickness function prescription,
$R^{Be/p}_{ATF}$,
specifically,
\beq
100\frac{R^{Be/p}_{GD}(p_T)-R^{Be/p}_{ATF}(p_T)}{R^{Be/p}_{GD}(p_T)}~.
\label{GD-ATF}
\eeq
As in the case of $^{184}W$, $R^{Be/p}_{UD}$ and $R^{Be/p}_{ATF}$ are
equal to about one percent or better.  The contribution of the
center-of-mass correction $\delta R^{Be/p}_{GD}$ to $R^{Be/p}_{GD}$,
specifically,
\beq
100\frac{\delta R^{Be/p}_{GD}(p_T)}{R^{Be/p}_{GD}(p_T)}
\label{cm-GD}
\eeq
is shown as the dashed curve in Fig.~\ref{delBef}.  The
center-of-mass correction for $^9Be$ is clearly quite small;
as for $^{184}W$, it contributes less than a percent to the nuclear
ratio.  Because the center-of-mass correction in Eq.~(\ref{cm}) is
very small even for light nuclei, it will
remain small for all realistic choices of density, and we therefore
neglect it in our final calculations.

The validity of the average $T_A$ prescription of Eq.~(\ref{avT1}) is
confirmed for light nuclei by observing that $U_0(r_T)$,
Eq.~(\ref{main}), is preserved under this substitution up to second
order in $T_A$ in its multiple scattering expansion. This observation
explains why the uniform density and the gaussian density models work
essentially equally as well for $^9Be$.  It is perhaps surprising that
the average thickness function prescription works so well even for
heavy nuclei, when the higher order multiple scattering terms are
important. The result is however intuitive, and it shows that on the
average the quark propagates through half the nucleus before undergoing
a Drell-Yan reaction. Based on our comparisons of the average thickness
function approximation to numerical calculations in models, we conclude
that the average thickness function approximation is justified for
applications to data, and we will use it in the main body of the paper
for calculating $\sigma^{pA}_{DY}$.

\end{document}